\documentclass[twoside]{amsart}

\usepackage{amssymb}
\usepackage{epsfig,epic,eepic,rotate,graphics,fancybox}
\usepackage[usenames,dvipsnames]{color}
\usepackage{graphicx}
\usepackage{epstopdf}

\usepackage[sc]{mathpazo} 
\usepackage[T1]{fontenc} 
\usepackage{microtype} 

\usepackage[left=20mm,top=32mm,columnsep=20pt]{geometry} 
\usepackage{multicol} 
\usepackage{booktabs} 
\usepackage{float} 
\usepackage{hyperref} 

\usepackage{lettrine} 
\usepackage{paralist} 

\usepackage{abstract} 

\usepackage{titlesec} 
\renewcommand\thesection{\Roman{section}} 
\renewcommand\thesubsection{\Roman{subsection}} 
\titleformat{\section}[block]{\large\scshape\centering}{\thesection.}{1em}{} 
\titleformat{\subsection}[block]{\large}{\thesubsection.}{1em}{} 

\def\bm#1{\mbox{\boldmath $#1$}}

\title{\Large{Mixtures of Hard Ellipsoids and Spheres : stability of the nematic phase}} 

\author[C.E. Alvarez]{Carlos E. Alvarez} \author[M. Mazars]{Martial Mazars}
\address[C.E. Alvarez]{Departamento de Matem\'aticas, Universidad del Rosario, Calle 14 No. 6-25, Bogot\'a, COLOMBIA.}
\email[C.E. Alvarez]{alvarezca.carlos@urosario.edu.co}
\address[M. Mazars]{Laboratoire de Physique Th\'eorique (UMR 8627), Universit\'e de Paris-Sud and CNRS, B\^atiment 210, 91405 Orsay Cedex, FRANCE.}
\email[M. Mazars]{Martial.Mazars@th.u-psud.fr}
\date{}

\begin{document}

\maketitle 



\begin{abstract}
The stability of liquid crystal phases in presence of small amount of non-mesogenic impurities is of general interest for a large spectrum of technological applications and in the theories of binary mixtures. Starting from the known phase diagram of the hard ellipsoids systems, we propose a simple model and method to explore the stability of the nematic phase in presence of small impurities represented by hard spheres. The study is performed in the isobaric ensemble with Monte Carlo simulations.
\end{abstract}


\begin{multicols}{2} 

\section{Introduction}

Fluids containing elongated particles may present a phase transition from an isotropic phase to liquid crystal mesophase \cite{Chandrasekhar:92,deGennes:93,Care:05}, in which orientational order  is present while there is only partial (smectic or columnar phases) or no translational order (nematic phase). A simple theoretical model for the isotropic-nematic transition was described in the Onsager's theory \cite{Onsager:49} of long cylindrical particles. In this model, the isotropic-nematic transition is entropy-driven by the steric interactions.\\
After the first Monte Carlo computations on two dimensional hard ellipses \cite{Vieillard:69,Vieillard:72} and three dimensional spherocylinders\cite{Vieillard:74}, several hard core models of mesogenic molecules have been developed and used in computer simulations \cite{Care:05} and in density functional calculations \cite{Poniew:92,Hartmut:99,Cheung:04}. These studies have permitted to obtain a better understanding of the liquid crystal phases and their relations with the shape of particles. The hard spherocylinder and spheroid models are certainly among the most studied systems because of their simplicity.\\  
The phase diagram for hard spheroids, both prolate and oblate, has been obtained  \cite{Frenkel:85,Mulder:85,Camp:96a,Pfleiderer:07} for different aspect ratios.  The phase diagram for prolate spheroids, depending on the density and aspect ratio, has five different phases : isotropic, nematic and three crystal phases \cite{Pfleiderer:07} ; a plastic solid, a fcc crystal and a simple monoclinic lattice with a basis of two ellipsoids (SM2 crystal).\\
Additionally to the study of the properties of monodisperse systems of rod-like or disk-like particles, there has been an increasing interest in the behavior of mixtures of particles with different geometry  \cite{Antypov:03,Varga:05,Konig:06,Kleshchanok:12,Green:12,Heras:13}. The properties of these mixtures is of technological interest as the presence of impurities in liquid crystal materials affects the orientational order and the electrical properties of compounds \cite{Ohgawara:81,Costa:01,Hung:12}. For instance, when the mole fraction of non-mesogenic cyclohexane molecules increases in a liquid crystal phase of octylcyanobiphenyls, one observes a decrease of the nematic-isotropic transition temperature, making the nematic phase less stable \cite{Denolf:07}. Obviously, the stability of the mesogenic phases in presence of impurities is strongly dependent on the nature (steric, electrostatic, etc.) of the intermolecular interactions between the impurities and the mesogenic molecules. The present work focus on steric interactions.\\       
For mixtures of hard rods and spheres experimental studies have been carried out \cite{Adams:98,Dogic:06} along with theoretical \cite{Sambroski:94,Martinez:06} and simulation counterparts \cite{Bolhuis:94,Antypov:03}. In these studies, a demixing into a sphere rich phase and a rod rich phase may occur as the density and molar fraction of spheres increases ; the rod phase is in a liquid crystal  state \cite{Sambroski:94}. In the case of cylindrical rods it has been observed that the presence of the small spheres, after demixing, stabilizes the smectic phase with the spheres located between the layers of rods \cite{Martinez:06}. For hard Gaussian overlap particles at constant volume, which are approximations to prolate spheroids, an increase in the molar  fraction of spheres destabilizes the nematic phase, postponing it for larger volume fractions \cite{Antypov:03} (the sphere diameter is equal to the minor axis of the spheroid) and for a sphere molar fraction of $0.5$ a demixing of phases is observed.\\
In this letter, using Monte Carlo simulations, we present an exploratory analysis of the influence of the addition of small spherical impurities in the nematic phase of prolate spheroidal particles. We use relatively small spheres to fit better into the interstices between spheroids, in an attempt to postpone the demixing observed in such systems to larger densities and look instead at the effects on the stability of the nematic phase of the spheroids for two different aspect ratios. The simulations are done at constant pressure, that makes easier the location of the mixed system in the phase diagram of the pure system and then extract the influence of the impurity molar fraction on the stability of the nematic phase. 


\section{Model and numerical methods}

\begin{figure*}
\begin{minipage}{0.45\linewidth} 
  \begin{minipage}{\linewidth}
    \epsfig{file=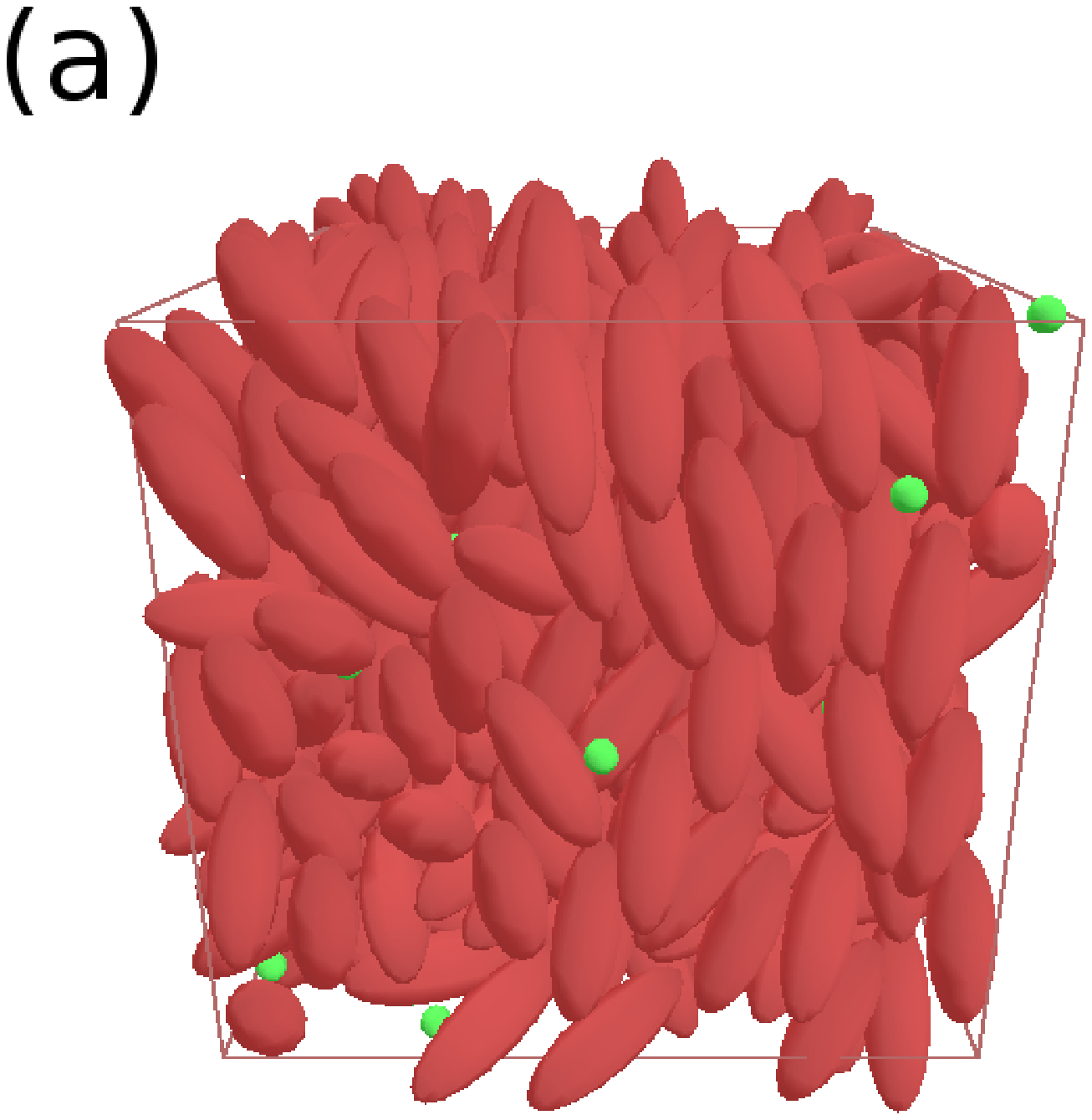,width=6.5cm,angle=0}
  \end{minipage}
  \vspace{0cm}
  \begin{minipage}{\linewidth}
    \epsfig{file=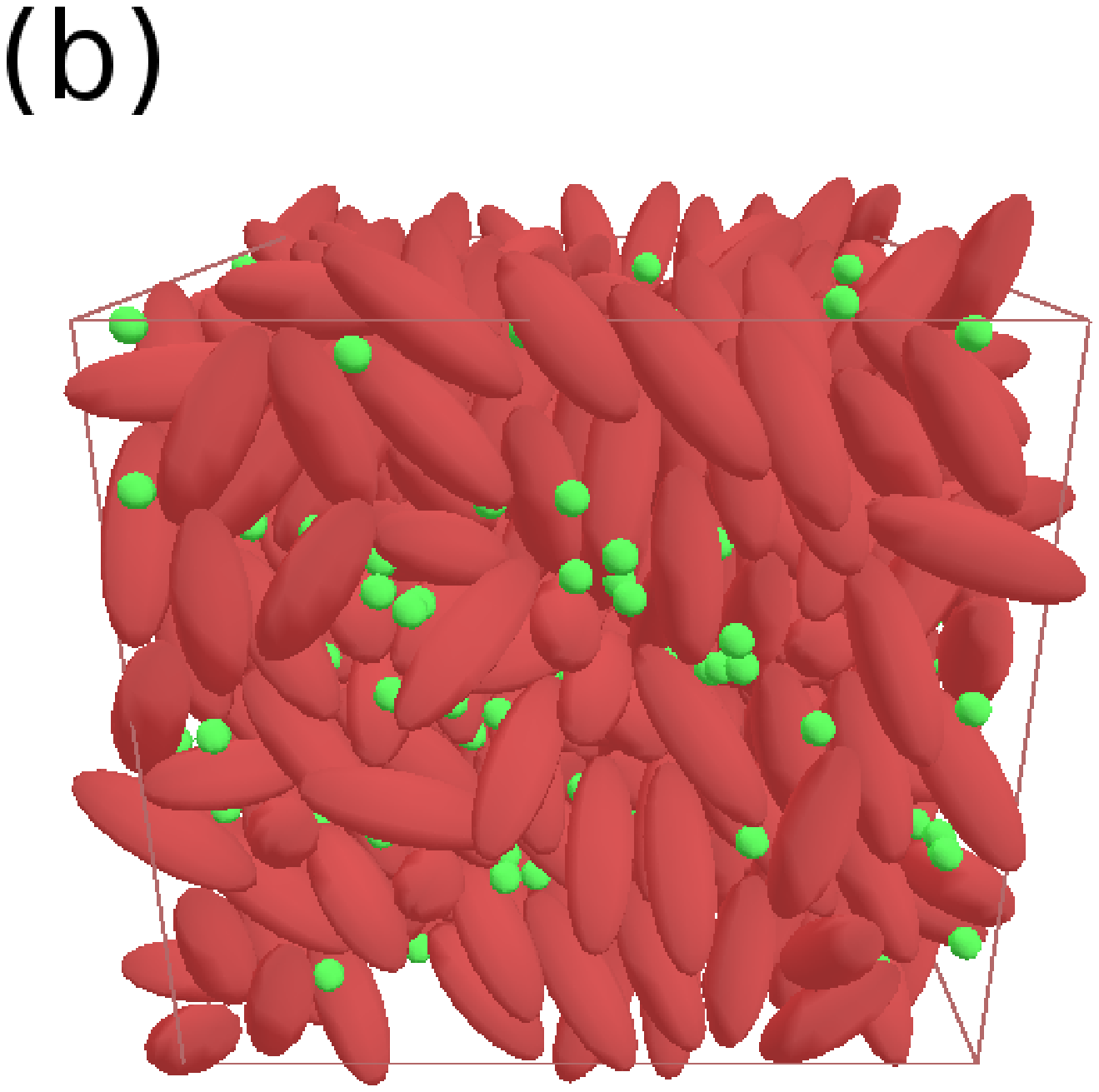,width=6.5cm,angle=0}
  \end{minipage}
  \vspace{0cm}
  \begin{minipage}{\linewidth}
    \epsfig{file=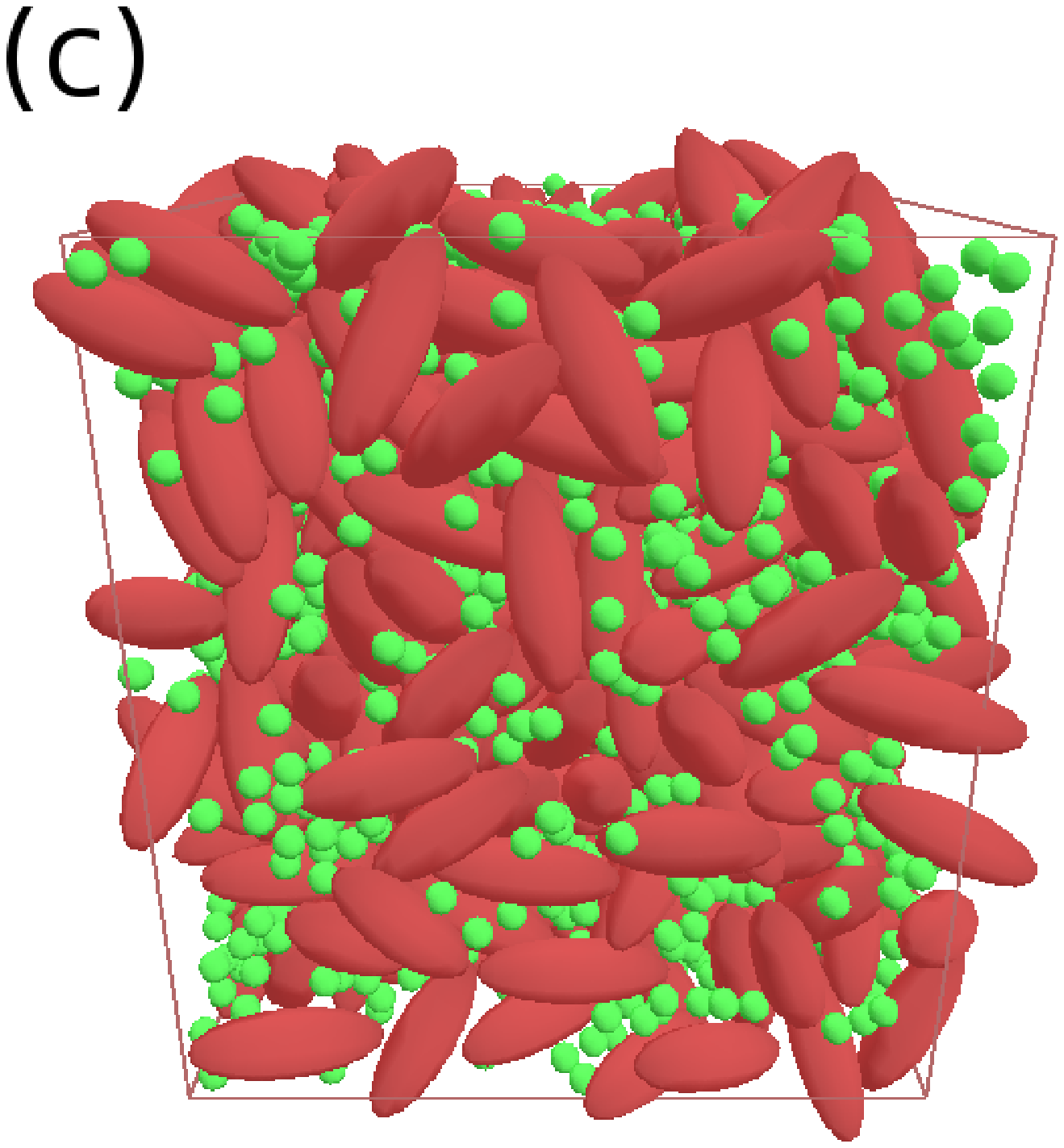,width=6.5cm,angle=0}
  \end{minipage}
\end{minipage}
\hspace{0cm}
\begin{minipage}{0.45\linewidth} 
  \begin{minipage}{\linewidth}
    \epsfig{file=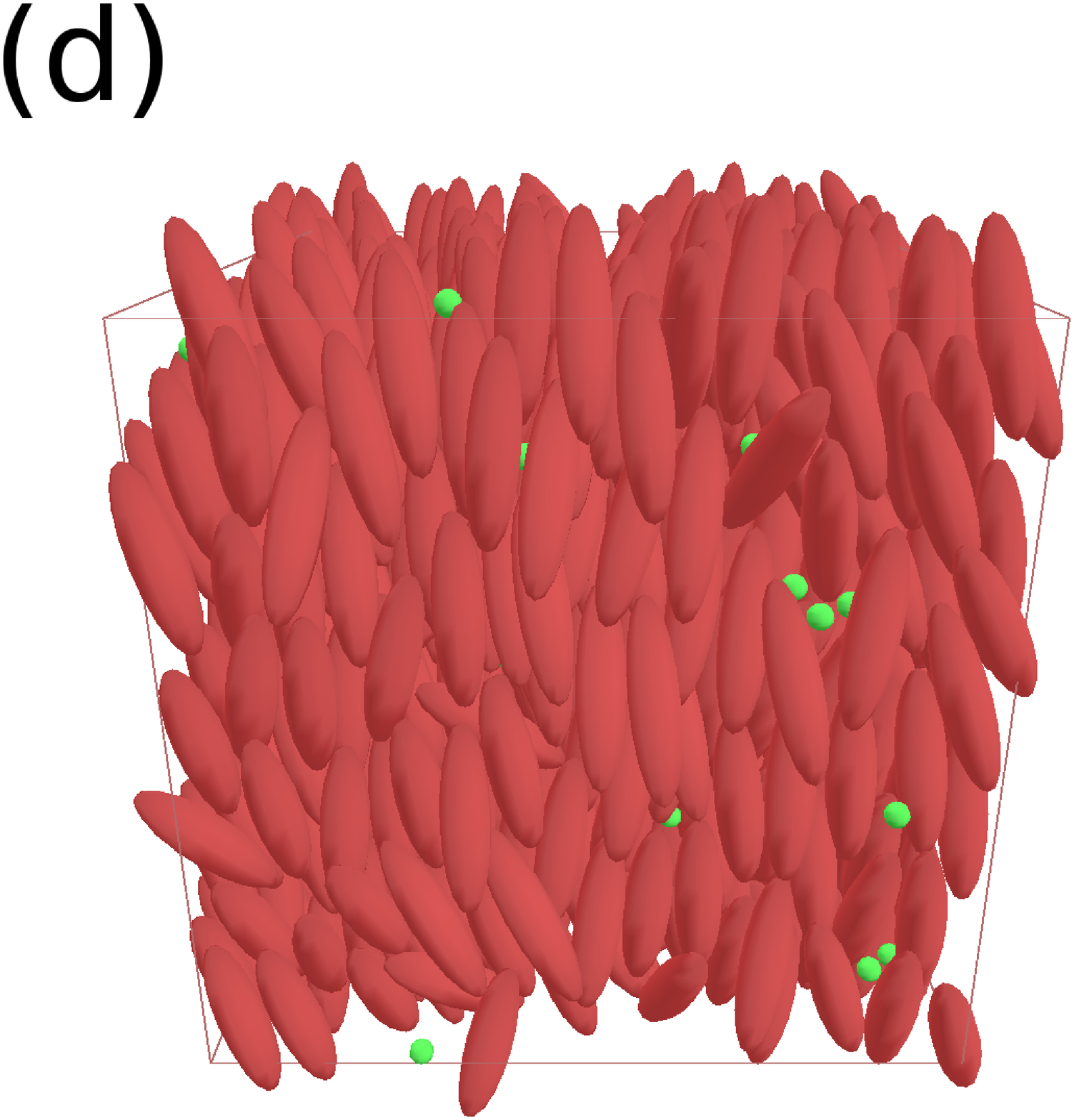,width=6.5cm,angle=0}
  \end{minipage}
  \vspace{0cm}
  \begin{minipage}{\linewidth}
    \epsfig{file=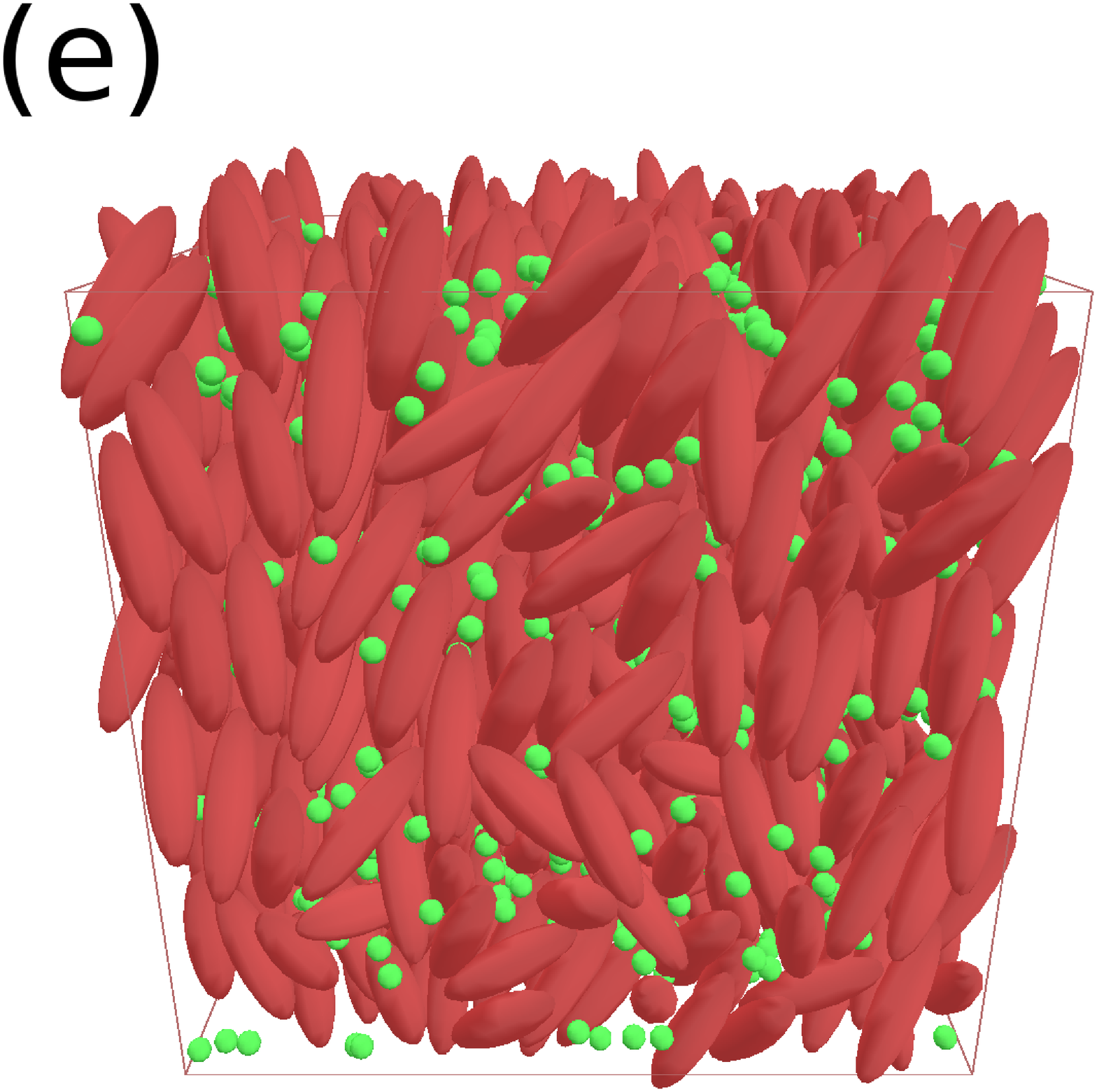,width=6.5cm,angle=0}
  \end{minipage}
  \vspace{0cm}
  \begin{minipage}{\linewidth}
    \epsfig{file=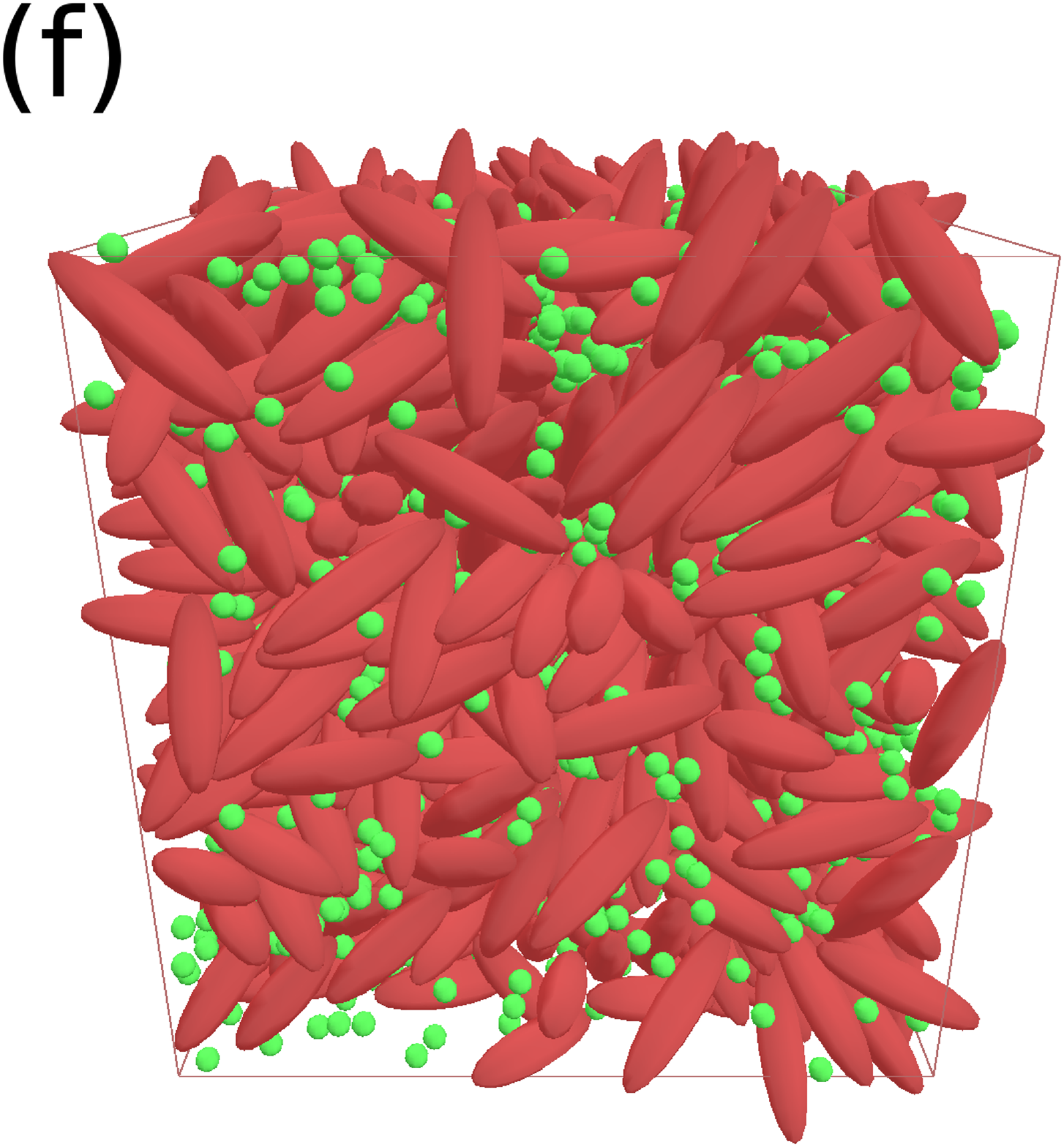,width=6.5cm,angle=0}
  \end{minipage}
\end{minipage}
\caption{\footnotesize {\bf Left:} Configurations for the system spheroids of aspect ratio $R/R'=3$ at  (a) $x_e=0.910$ ($G_2=0.56(3)$),(b) $x_e=0.625$ ($G_2=0.27(9)$), and (c) $x_e=0.250$ ($G_2=0.07(2)$).  {\bf Right:} Configurations for the system spheroids of aspect ratio $R/R'=4$  at (d) $x_e=0.910$  ($G_2=0.828(8)$),(e) $x_e=0.455$ ($G_2=0.46(2)$), and (f) $x_e=0.333$ ($G_2=0.08(2)$).}
\label{configs}
\end{figure*}

The Monte Carlo simulations are performed in the isobaric ($NPT$) ensemble, in which the pressure $P$ is chosen such that the pure system is located in the nematic phase, as specified with the known phase diagram of the monodisperse system of prolate spheroids \cite{Frenkel:85,Pfleiderer:07}. In order to estimate the loss of stability of the nematic phase in presence of small impurities, the isobaric ensemble should be preferred to the canonical ($NVT$) ensemble, in which the fixed volume may trap the system in a finite region of the phase space or lead to some unphysical demixing.\\
Two spheroid-sphere mixtures, with spheroids of two different aspect ratios (3 and 4) are considered in this work ; the number of spheroids is noted $N_e$ and $N_s$ is the number of impurities (hard spheres). Let $R$  and  $R'$ be respectively the major and minor semiaxes of the spheroid, the aspect ratio is defined by $R/R'$ ;  the diameter $\sigma$ of the small hard spheres is chosen equal to the minor semi-axis $R'$ of the spheriods with an aspect ratio 4. The volume of a prolate spheriod is $V_e=4\pi RR'^2/3$ ; with the choice of $\sigma$ for the aspect ratio 4 mixture, the volume of the small hard sphere is $V_s=R'V_e/(8R)=V_e/32$. For the mixtures with spheroids of aspect ratio 3, we have chosen to keep the volume ratio $V_s/V_e=1/32$ constant, such as the total excluded volume due to the sphere is in the same proportion in both mixed systems. The important parameters in the model are the shape ($R/R'$) and the molar fraction of spheroids ($x_e=N_e/(N_e+N_s)$).\\
\begin{figure}[H]
\centerline{\includegraphics[width=2.5in]{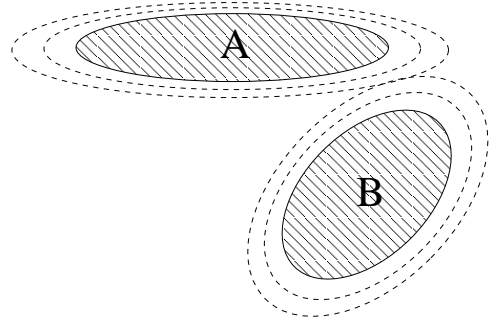}}
\caption{{\bf} Scaling factor $\mu$ to obtain tangent ellipsoids.}
\label{FigA1}
\end{figure}  
The reduced units used in this work are: the density of spheroids $\rho_e^*=8 R R'^2  (N_e/V) = 8 R R'^2 \rho_e$ with $V$ the volume of the simulation box and the pressure defined by  $P^*=8 R R'^2 \beta P$ with $\beta=1/k_B T$ ; in hard sphere systems ($x_e=0$) the reduced pressure is defined as  $p^*=\beta P \sigma^3$. For convenience of notations, we drop the asterisks in the following.\\ 
Before starting equilibration, the initial conditions for the systems were set by compressing them at a high pressure ($P=30$ for $R/R'=3$ and $P=25$ for $R/R'=4$), while maintaining the orientation of spheroids  fixed in the $z$ direction. From these initial configurations the orientation of the spheroids was freed and the systems were allowed to equilibrate.\\ 
From the initial conditions, $5\times10^5$ equilibration MC-steps  are used to equilibrate the systems, then all averages quantities are reset to zero and $2.5\times10^5$ additional MC-steps are used to compute averages (1 MC-steps consist of $N$ trial moves of particles, plus one trial change of the volume of the simulation box).\\ 
For mixtures with aspect ratio 4 spheroids, most of the computations have been done with $N_e=1000$ spheroids and with a number of spheres between $0$  and $2000$. For the larger molar fractions of  spheres, to keep the computation time reasonable, we have decreased the number of spheroids ($200\leq  N_e\leq 1000$). For all computations with this aspect ratio, the pressure is fixed at $P=20$.\\
The computations for mixtures with aspect ratio 3 spheroids are done with $N_e=500$ and $N_s$ between 0  and 1500 ; for these systems, the pressure is chosen at  $P=25$.\\
Subsequently, the molar fraction of small spheres, defined as  $x_s=N_s/(N_e+N_s)$, is increased to observe the influence of the concentration of small spheres on the stability of the nematic phase of the spheroids.\\ 
The correspondence between the reduced pressure $p$ for hard sphere systems and the reduced pressure $P$ for spheroids is given by 
\begin{equation}
\displaystyle p=\frac{R'}{8R}P
\label{pHSphere}
\end{equation}
Since the coexistence pressure for the fluid-solid first order transition for the hard sphere systems is $p_{liq/sol}\simeq 11.6$ \cite{Binder:10}, with the thermodynamical parameters chosen in our study, the pure hard sphere system $x_s=1$ ($x_e=0$) is in the fluid phase for $P=20$ (i.e. $p=5/8$ for systems with aspect ratio 4) and for $R/R' = 3$ with $P=25$. With these choices for the pressure of the mixtures (aspect ratio 3 and 4), the pure systems($x_s\rightarrow 0$) are in a nematic phase and the state for the infinite dilution of spheroids ($x_s\rightarrow 1$) is the fluid of hard spheres.\\
\begin{figure}[H]
\epsfig{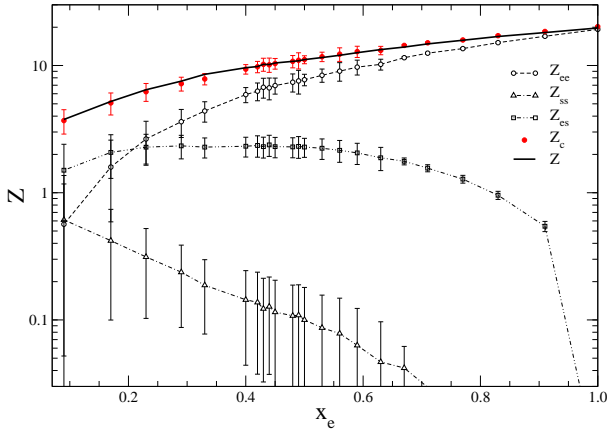}
\caption{\footnotesize Partial compressibility factors for systems with an aspect ratio $R/R'=4$. $Z$ is the imposed compressibility factor in the isobaric ensemble and $Z_c$ is the computed compressibility factor. $Z_{ee}$, $Z_{ss}$ and $Z_{es}$ are respectively the spheroid-spheroid, sphere-sphere and spheroid-sphere contributions to the compressibility.}
\label{Comp4}
\end{figure}
To find overlaps between the particles, we use the contact function derived earlier in refs.\cite{Vieillard:72,Perram:85} as described below.\\ 
The orientation of an ellipsoid $A$ and its shape are defined by three vectors $\bm{R}_i$ ($i=1,2,3$) whose direction and magnitude are those of the semiaxes of the ellipsoid. The contact function is defined as
\begin{equation}
\displaystyle F(\bm{r}_{AB},\Omega_A,\Omega_B)=\mbox{max}_{0\leq\lambda\leq 1}\left(\lambda(1-\lambda)\bm{r}^{T}_{AB}\bm{C}(\lambda)\bm{r}_{AB} \right)
\label{Contact}
\end{equation}
If $F(\bm{r}_{AB},\Omega_A,\Omega_B) < 1$, then overlap is found and if $F(\bm{r}_{AB},\Omega_A,\Omega_B) = 1$ the ellipsoids are tangent.\\
In Eq.(\ref{Contact}), the matrix $\bm{C}(\lambda)$ is defined by 
\begin{equation}
\displaystyle \bm{C}(\lambda) = \left[\lambda \bm{B}+(1-\lambda)\bm{A}\right]^{-1}
\label{matrixC}
\end{equation}
where the two matrices $\bm{A}(\Omega_A)$ and $\bm{B}(\Omega_B)$ are defined from the cartesian orientation for each ellipsoid as
\begin{equation}
\displaystyle \bm{A}(\Omega_A)=\sum_{i=1}^3\bm{R}_i(\Omega_A)\bm{R}_i^{T}(\Omega_A)
\end{equation}
and same definition for the ellipsoid $B$. On figure \ref{configs}, we show some snapshots of the spheroid-sphere mixtures for aspect ratios 3 and 4, at different molar fraction of impurities.\\
Since, for all $\lambda$, the matrix $\bm{C}(\lambda)$ is positive definite, we may write 
\begin{equation}
\displaystyle F(\bm{r}_{AB},\Omega_A,\Omega_B)=\mu^2
\label{mu2}
\end{equation}
where $\mu$ is a scaling factor that produces two tangent ellipsoids $A$ and $B$ (cf. Fig\ref{FigA1}) ; $\lambda_m$ is the value found in the maxima of Eq.(\ref{Contact}).\\ 
Up to second order in density, the pressure in the systems can be computed with the method proposed by Perram and Wertheim 
\cite{Perram:85}, as
\begin{equation}
\label{Fpres}
\beta P_c=\rho+\frac{1}{3V}\left<\sum_{\{kl\},k<l}2F(\bm{r}_{kl})\ \delta\left(F(\bm{r}_{kl})-1\right)\right>,
\end{equation}
where $P_c$ is the pressure, $\beta=1/k_BT$, $V$ is the volume and $F(\bm{r})$ is the contact function between two particles (spheroids and/or sphere).\\
The configurational integral for a system of $N$ hard convex particles in a volume $V$ is 
\begin{equation}
\begin{array}{ll}
\displaystyle Q(V,N)=\frac{1}{(8\pi V)^N N ! }&\\
&\\
\displaystyle\times \int\cdots\int d^N\bm{r}d^N\Omega \prod_{\{ij\}, i<j}\Theta\left(F(\bm{r}_{ij},\Omega_i,\Omega_j)-1\right) &
\label{zzz}
\end{array}
\end{equation}
where $\Theta(x)$ is the Heaviside function. From the thermodynamic relation
\begin{equation}
\displaystyle \beta P=\frac{\partial \ln Z}{\partial V}=\frac{1}{Z}\frac{\partial Z}{\partial V},
\label{ppp}
\end{equation}
and the functional relation 
\begin{equation}
\begin{array}{ll}
\displaystyle\frac{\partial }{\partial V} \prod_{\{ij\}, i<j}\Theta\left(F(\bm{r}_{ij},\Omega_i,\Omega_j)-1\right)&\\
&\\
\displaystyle \displaystyle=\left[\sum_{\{kl\}, k<l}F(\bm{r}_{kl})\delta(F(\bm{r}_{kl})-1) \right]&\\ 
&\\
\displaystyle \times\prod_{\{ij\}, i<j}\Theta\left(F(\bm{r}_{ij},\Omega_i,\Omega_j)-1\right)&
\end{array}
\end{equation}
Eq.(\ref{Fpres}) follows.\\
The difficulty in the computation of pressure as an average with Eq.(\ref{Fpres}) stems from the Dirac distribution due to the hard core interactions. In Monte Carlo computations, the probability of obtaining configurations with tangent ellipsoids is too small for a simple computation of the pressure with Eq.(\ref{Fpres}) as an average over the MC trajectory in the phase space. To overcome this difficulty, we count the number of particles that overlap when a scaling factor $\mu^2(>1)$  is applied to all particles in the system (cf. Fig.\ref{FigA1}). Then, for each value of $\mu^2$ chosen, we compute the pressure according Eq.(\ref{Fpres}), this provides $P_c(\mu^2)$ ; the measurement of the pressure $P_c$ for a given configuration follows from an extrapolation of $P_c(\mu^2)$ at contact ($\mu^2=1$).\\  
These computations are useful to check the consistency of the $NPT$ computations and it also permits an analysis of results with the partial components of the virial.\\
To compute the partial virials, the sum over pairs in Eq.(\ref{Fpres}) is split into three contributions as
\begin{equation}
P_{a b}\displaystyle=\frac{1}{3V}\left<\sum_{\{k_{a}l_{b}\}}\mbox{}'\mbox{ }2F(\bm{r}_{k_{a}l_{b}})\ \delta\left(F(\bm{r}_{k_{a}l_{b}})-1\right)\right>
\label{partP}
\end{equation}
where the index $k_a$ runs over the spheroids if $a\equiv e$ or spheres if $a\equiv s$ (same notations for $l_b$) ; the prime in the summation indicates that the contributions $k_a=l_b$ are excluded if  $a\equiv b$. More precisely, in equation (\ref{partP}), $P_{ee}$ is the partial virial due to spheroid-spheroid contact, $P_{ss}$ is the partial virial due to sphere-sphere contact, and $P_{es}$ is the partial virial due to spheroid-sphere contact.\\
From the definition of partial virials, we define the partial compressibility factors as 
\begin{eqnarray}
Z_c=\frac{\beta P_c}{\rho_c}&=&1+\frac{1}{\rho_c k_B T}\left(P_{ee}+ P_{es}+ P_{ss}\right)\nonumber\\
&&\nonumber\\
&=&1+Z_{ee}+Z_{es}+Z_{ss},
\label{partZ}
\end{eqnarray}
where $\rho_c$ is the average density obtained in the simulations.\\
A check of the consistency of the Monte Carlo computations is achieved by comparing the applied compressibility factor $Z=\beta P/\rho$ in the isobaric ensemble with the computed compressibility factor from Eq.(\ref{partZ}). This comparison is shown on Figure \ref{Comp4} for systems with aspect ratio 4 ; as it is explicitly shown in the figure, the agreement is excellent. On this figure, we report also the partial compressibility factors, as defined in equation (\ref{partZ}).


\section{Mechanical stability of the nematic phase in presence of spherical impurities}

To measure the nematic order in the system we use the order parameter $G_2$, which is the largest  eigenvalue of the matrix
\begin{equation}
Q_{\alpha\beta}=\frac{1}{2N_e}\sum_{i=1}^{N_e}\left(3\mu_{i\alpha}\mu_{i\beta}-\delta_{\alpha\beta}\right),
\end{equation}
\begin{figure}[H]
\epsfig{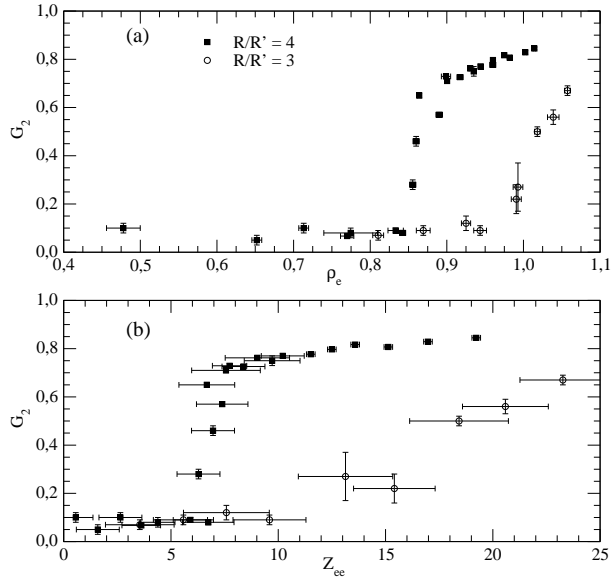}
\caption{\footnotesize Nematic Order parameter $G_2$ for mixtures of spheroids and hard sphere for systems with aspect ratio 4 and 3 as a function of the spheroid density (a) and spheroid-spheroid partial compressibility factor (b). }
\label{Gplots}
\end{figure} 
where $i$ denotes the particle and the indexes $\alpha$ and $\beta$ label the cartesian components of the unit vector $\hat{\bm{\mu}}$, parallel to the revolution axis of the spheroid.\\
In Figure \ref{Gplots}, we represent the change of the order parameter $G_2$ against the spheroid density $\rho_e$ and spheroid-spheroid partial compressibility factor $Z_{ee}$. As the molar fraction of spheroids $x_e$ increases, both $\rho_e$ and $Z_{ee}$ increase and the system undergoes a phase transition from the isotropic fluid to the nematic phase.\\
From the phase diagram\cite{Frenkel:85,Mulder:85,Pfleiderer:07} of hard spheroid systems ($x_e=1$),  the nematic phase for aspect ratio $R/R'=4$ is stable for particle density  $0.7\lesssim\rho_e\lesssim1.04$, and for an aspect ratio of 3, the region of stability of the nematic phase is narrower ($1.0\lesssim\rho_e\lesssim1.1$).\\
For the mixtures studied in this work, we locate roughly the isotropic-nematic transition at $x_e(I/N)\simeq 0.45$ and we compute $\rho_e(I/N)\simeq 0.87$ and $P_{ee}(I/N)\simeq 13$ for systems with aspect ratio 4. For $R/R' = 3$, the transition to the nematic phase is less displaced: we found $x_e(I/N)\simeq 0.83$, $\rho_e(I/N)\simeq 1.04$ and $P_{ee}(I/N)\simeq 22$, but the order is also less marked (Fig.\ref{Gplots}) suggesting that the isotropic-nematic coexistence is larger than that of pure systems.\\
In previous simulations, done on pure spheroid systems ($x_e=1$), the nematic-isotropic transition has been located at $\rho_{I/N}\simeq 1.0$, with a coexistence pressure $P\simeq 18.7$, for spheroids with aspect ratio 3 and at $\rho_{I/N}\simeq 0.7$ for aspect ratio 4 (see for instance refs. \cite{Frenkel:85,Mulder:85,Pfleiderer:07}). The locations of the isotropic-nematic transitions in the mixtures show that the presence of impurities destabilizes the nematic phase by increasing the spheroid density at which the transition occurs ; a larger spheroid-spheroid contribution to the virial is also needed to obtain a significant nematic order.\\
Figure \ref{configs} shows the configurations of the system in the nematic and isotropic phases, and close to the phase transition for the $R/R'=3$ and $R/R'=4$ cases.\\ 
It is possible also that at high enough molar fraction of spheres, a demixing between hard spheres and spheroids occurs, and then a reentrance of the nematic phase in the  spheroid-rich regions, similarly  as it was observed in previous studies \cite{Antypov:03,Sambroski:94}. However, with the parameters chosen in the present work, the size of the sample and within the length of our simulations, we did not observe any signs of demixing.


\section{Discussion}

Since most the real substances are not pure, the phase diagrams and the thermodynamic properties of mixtures are of great importance for both technological and experimental applications.\\ The binary phase diagrams for the liquid gas transition are various and strongly dependent on the identity of the components of the mixture, at least there exist six main classes of binary phase diagrams with subclasses in each principal classes (see for instance chapter 7 of ref.\cite{Gray:11}). When solid regions of the phase diagrams are added, the complexity of the binary phase diagrams may increase significantly \cite{Streett:76}.\\
In this communication, we have presented a simple model of binary system and a method to study easily the stability of the nematic phase in presence of small impurities. This binary mixture of hard spheroids and spheres has two thermodynamical parameters : the pressure $P$ and the impurity molar fraction $x_s$ ; and two geometrical parameters : the aspect ratio $R/R'$ of the spheroids and the volume ratio $V_s/V_e$. At fixed geometrical parameters and pressure, we have shown that the isotropic-nematic transition can be studied quite easily. A longer study would explore the full ($P$, $x_s$)-region for various aspect ratio, at a fixed volume ratios ; knowing that phase diagrams at $x_s=0$ is the phase diagram of the pure spheroid systems \cite{Pfleiderer:07} and at $x_s=1$, for $V_s/V_e=1/32$ because of Eq.(\ref{pHSphere}), the system is a gas of hard sphere.\\
For the two aspect ratios considered in the present work, the increase of the impurities molar fraction renders the nematic phase less stable. There are some indications that as the aspect ratio of the spheroids decreases, the coexistence region between the isotropic and nematic phases becomes larger.


\bibliographystyle{unsrt} 
\bibliography{Alvarez_Mazars_MHE} 

\begin{thebibliography}{10}

\bibitem{Chandrasekhar:92}
S.~Chandrasekhar.
\newblock {\em Liquid Crystals - Second edition}.
\newblock Cambridge University Press, Cambridge, 1992.

\bibitem{deGennes:93}
P.G. de~Gennes and J.~Prost.
\newblock {\em The Physics of Liquid Crystals - Second edition}.
\newblock Oxford University Press, Oxford, 1993.

\bibitem{Care:05}
C.~M. Care and D.~J. Cleaver.
\newblock {\em Rep. Prog. Phys.}, 68:2665, 2007.

\bibitem{Onsager:49}
L.~Onsager.
\newblock {\em Ann. N. Y. Acad. Sci.}, 51:627, 1949.

\bibitem{Vieillard:69}
J.~Vieillard-Baron.
\newblock {\em J. Phys. (Paris) Colloq. (Suppl. C4)}, 30:22, 1969.

\bibitem{Vieillard:72}
J.~Vieillard-Baron.
\newblock {\em J. Chem. Phys.}, 56:4729, 1972.

\bibitem{Vieillard:74}
J.~Vieillard-Baron.
\newblock {\em Mol. Phys.}, 28:809, 1974.

\bibitem{Poniew:92}
A.~Poniewierski and T.~J. Cluckin.
\newblock {\em Mol. Cryst. Liq. Cryst.}, 212:61, 1992.

\bibitem{Hartmut:99}
G.~Hartmut and L.~Hartmut.
\newblock {\em J. Phys.: Condens. Matter}, 11:1435, 1999.

\bibitem{Cheung:04}
D.~L. Cheung and F.~Schmid.
\newblock {\em J. Chem. Phys.}, 120:9185, 2004.

\bibitem{Frenkel:85}
D.~Frenkel and B.~M. Mulder.
\newblock {\em Mol. Phys.}, 55:1171, 1985.

\bibitem{Mulder:85}
B.~M. Mulder and D.~Frenkel.
\newblock {\em Mol. Phys.}, 55:1193, 1985.

\bibitem{Camp:96a}
P.~J. Camp, C.~P. Mason, M.~P. Allen, A.~K. Anjali, and D.~A. Kofke.
\newblock {\em J. Chem. Phys.}, 105:2837, 1996.

\bibitem{Pfleiderer:07}
P.~Pfleiderer and T.~Schilling.
\newblock {\em Phys. Rev. E}, 75:020402, 2007.

\bibitem{Antypov:03}
D.~Antypov and D.~J. Cleaver.
\newblock {\em Chem. Phys. Lett.}, 377:311, 2003.

\bibitem{Varga:05}
S.~Varga, K.~Purdy, A.~Galindo, S.~Fraden, and G.~Jackson.
\newblock {\em Phys. Rev. E}, 72:051704, 2005.

\bibitem{Konig:06}
P.-M. K\"onig, R.~Roth, and S.~Dietrich.
\newblock {\em Phys. Rev. E}, 74:041404, 2006.

\bibitem{Kleshchanok:12}
D~Kleshchanok, J.~M. Meijer, A.~V. Petukhov, G.~Portale, and H.~N.~W.
  Lekkerkerker.
\newblock {\em Soft Matter}, 8:191, 2012.

\bibitem{Green:12}
M.~J. Green.
\newblock {\em J. Polym. Sci. Part B}, 50:1321, 2012.

\bibitem{Heras:13}
D.~de~las Heras and M.~Schmidt.
\newblock {\em Phil. Trans. R. Soc. A}, 371:20120259, 2013.

\bibitem{Ohgawara:81}
M.~Ohgawara and T.~Uchida.
\newblock {\em Jpn. J. Appl. Phys.}, 20:L75, 1981.

\bibitem{Costa:01}
M.~R. Costa and R.~A.~C. Altafim.
\newblock {\em Liq. Cryst.}, 28:1779, 2001.

\bibitem{Hung:12}
H-.Y. Hung, C-.W. Lu, C-.Y. Lee, C-.S. Hsu, and Y-.Z. Hsieh.
\newblock {\em Anal. Methods}, 4:3631, 2012.

\bibitem{Denolf:07}
K.~Denolf, G.~Cordoylannis, C.~Glorleux, and J.~Thoen.
\newblock {\em Phys. Rev. E}, 76:051702, 2007.

\bibitem{Adams:98}
M.~Adams, Z.~Dogic, S.~L. Keller, and S.~Fraden.
\newblock {\em Nature}, 393:349, 1998.

\bibitem{Dogic:06}
Z.~Dogic and S.~Fraden.
\newblock {\em Curr. Op. Coll. Int. Sci.}, 11:47, 2006.

\bibitem{Sambroski:94}
A.~Sambroski and G.~T. Evans.
\newblock {\em J. Chem. phys.}, 101:6005, 1994.

\bibitem{Martinez:06}
Y.~Mart\'{\i}nez-Rat\'on, G.~Cinacchi, E.~Velasco, and L.~Mederos.
\newblock {\em Eur. Phys. J. E}, 21:175, 2006.

\bibitem{Bolhuis:94}
P.~Bolhuis and D.~Frenkel.
\newblock {\em J. Chem. Phys.}, 101:9869, 1994.

\bibitem{Binder:10}
T.~Zykova-Timan, J.~Horbach, and K.~Binder.
\newblock {\em J. Chem. Phys.}, 133:014705, 2010.

\bibitem{Perram:85}
J.~W. Perram and M.~S. Wertheim.
\newblock {\em J. Comp. Phys.}, 58:409, 1985.

\bibitem{Gray:11}
C.G. Gray, K.E. Gubbins, and C.G.. Joslin.
\newblock {\em Theory of Molecular Fluids - Volume 2: Applications}.
\newblock Oxford University Press, Oxford, 2011.

\bibitem{Streett:76}
W.B. Streett.
\newblock {\em Icarus}, 29:173, 1976.

\end{thebibliography}


\end{multicols}

\end{document}